%% file: main.tex
\documentclass{article}
\usepackage{graphicx}
\usepackage{geometry}
\usepackage{natbib,times,tikz,fontawesome6,amsmath,amsfonts}
\usepackage[labelfont=bf]{caption}
\usepackage[hidelinks]{hyperref}

\renewcommand{\emph}[1]{\textit{#1}}

\title{\bf When legs and bodies synchronize:\\
Two-level collective dynamics in dense crowds}

\author{Thomas Chatagnon$^1$, Mohcine Chraibi$^{1,3}$, Julien Pettré$^2$\\ Armin Seyfried$^{1,3}$ and Antoine Tordeux$^4$\\[2mm]
\small
\begin{tabular}{l}
$^1$ Forschungszentrum Jülich, Institute for Advanced Simulation (IAS), Germany\\
$^2$ Inria Rennes and Universities of Rennes 1 and Rennes 2, France\\
$^3$ University of Wuppertal, Faculty of Architecture and Civil Engineering, Germany\\
$^4$ University of Wuppertal, Faculty of Mechanical Engineering and Safety Engineering, Germany
\end{tabular}}

\date{}

\begin{document}

\maketitle

\begin{abstract}
Ultra-dense crowds, in which physical contact between people cannot be avoided, pose major safety concerns. Nevertheless, the underlying dynamics driving their collective behaviours remain poorly understood. Existing dense crowd models, mostly two-dimensional and contact-based, overlook biomechanical mechanisms that govern individual balance motion. In this study, we introduce a minimal two-level pedestrian model that couples upper body and legs dynamics, allowing us to capture transitions between balanced and unbalanced states at the individual scale. Whereas previous models fail to achieve it, this coupling gives rise to emergent collective behaviours observed empirically, such as self-organized waves and large-scale rotational motion within the crowd. The model bridges basic individual biomechanical concepts and macroscopic flow dynamics, offering a new framework for modelling and understanding collective motions in ultra-dense crowds.\\[2mm]
\end{abstract}

\section{Introduction}

Ultra-dense crowds, i.e., crowds of 4 to 5~ped/m$^{2}$ and above where physical interactions between people cannot be avoided, are a major safety concern worldwide, with accidents reported every year \cite{feliciani2023trends}. Despite the potential hazards, the mechanisms underlying the collective dynamics of ultra-dense crowds remain poorly understood. As an example, most recent analysis of density wave and rotational movements in ultra-dense crowds could not reveal its possible causes~\cite{bottinelli2018can,gu2025emergence}. What in individual behaviour can explain such large, and dangerous, patterns? 
Early interpretations of crowd disasters commonly invoked the notion of \emph{panic}, suggesting that individuals in emergencies lose rational control and behave selfishly or chaotically \cite{lebon1896crowd,smelser1962theory,quarantelli1975panic}. 
Building on these theories, simulation models have consistently sought to incorporate the propagation of emotional states and their effects on individual motion and interactions with neighbours~\cite{helbing2002crowd,trivedi2018agent,xu2020emotion}.
However, numerous empirical studies and post-event analyses have since disproven this view \cite{Quarantelli2001,drury2009everyone,haghani2019panic,sieben2023inside}. 
Eyewitness accounts, video evidence, interviews and sociological investigations consistently show that people in extreme crowding conditions generally remain calm, cooperative, and even altruistic.
Fatal outcomes may arise not from irrational behaviour, but from collective mechanical instabilities -- when local pressures, force chains, and density waves propagate through the mass of pedestrians, leading to loss of balance, fainting, and even asphyxia \cite{sieben2023inside}. 

This paradigm shift, integrating physical modelling with behavioural realism, reframes dense crowds as complex dynamical systems governed by biomechanical interactions rather than panic. Existing models, which primarily describe the passive transmission of motion between individuals through contact interactions and basic particle-like force exchanges, are not designed to capture additional active mechanisms. In particular, balance control behaviours can significantly influence the contact propagation dynamics, for example by amplifying, attenuating, or redirecting transmitted perturbations \cite{Feldmann_Adrian_Boltes_2024}. These active control processes constitute plausible explanatory factors for empirically observed phenomena that cannot be accounted for by passive propagation alone. 
Indeed, empirical studies have revealed a rich variety of collective patterns in dense pedestrian gatherings. Pioneering observations in Mecca, Saudi Arabia, during the Hajj in 2006 identified stick-slip instabilities and turbulent-like flows, sometimes referred to as earthquake-like \textit{crowd turbulence}, where high-density pedestrian motions lead to sudden, large-scale ruptures \cite{helbing2007dynamics}. 
Similar wave-like motions have been documented in large concert audiences, notably at the Oasis concert in 2005 in the United Kingdom \cite{bottinelli2018can}, where propagating density waves emerge spontaneously. 
Collective chiral oscillations have been observed in the Saint Fermín festival and the Love Parade in Duisburg, highlighting complex rotational and oscillatory modes in ultra-dense crowds \cite{gu2025emergence}. 
These studies underline that ultra-dense crowds are far from static; even under extreme compression, they display rich self-organized collective behaviour that occur without clear triggers and require mechanistic understanding.
Current models fail to predict how individual interactions translate into macroscopic crowd behaviour \cite{Lee_2005,Zhou_2017}. 
Understanding and modelling the collective dynamics of ultra-dense crowds is therefore critical not only for safety management but also for designing interpretable models capable of predicting and mitigating risky scenarios.

Classical microscopic pedestrian models, designed for low and intermediate densities at operational level, rely on concepts such as the fundamental diagram, social forces, collision avoidance, or trajectory optimization \cite{duives2013state,chraibi2019modelling,van2021algorithms}. 
These approaches have successfully described crowd behaviour under free and congested conditions, capturing flow, lane formation, and individual avoidance \cite{boltes2019empirical}. In these regimes, balance is maintained effortlessly through normal locomotion. However, physical contacts between pedestrians are continuous in ultra-dense situations and the dynamics changes fundamentally \cite{gu2025emergence,chatagnon2025exploring}.  
The contact forces can challenge postural control and standing balance must be actively maintained.  
Due to bio-mechanical constraints, the upper body and the legs can become constrained in different ways, preventing natural coordination and forcing pedestrians into potentially unbalanced postures. 
At such densities, pedestrian motion is primarily governed by physical forces and the bio-mechanical body response \cite{helbing2007dynamics,bottinelli2016emergent,wang2019modeling,gu2025emergence}, rather than intentional navigation behaviour and locomotion-based balance maintaining: classical modelling framework become inadequate.

In the literature, the representation of physical interactions in dense crowds relies on two tightly coupled components: the spatial representation of the crowd and the modelling of interaction effort \cite{chatagnon2025exploring}. 
At the largest scale, macroscopic approaches treat the crowd as a continuum and model focus on the overall dynamics. Such formulations have recently been used to capture rotational chiral oscillations without resolving individual interactions at the local level \cite{gu2025emergence}.
Hybrid micro–macro approaches based on kernel continuum formulations can been seen as a bridge between individual-level dynamics and macroscopic flows, offering a clearer separation between social and physical interactions while capturing density-dependent patterns and flow heterogeneities \cite{narain2009aggregate,golas2014continuum,van2021sph}.
Microscopic pedestrian models for dense crowds include individual interactions arising from body contact, such as normal repulsion and tangential friction, modelled by quadratic and higher-order nonlinear potentials \cite{helbing2000simulating,yu2007modeling}. 
For instance, the repulsive force is hyperbolic in the \emph{asocial} pedestrian model \cite{bottinelli2016emergent}. 
In \cite{silverberg2013collective}, the authors employ a Vicsek-like model with algebraic body contact and passive/active pedestrians to reproduce the dense crowd behaviour observed during concerts. Another example is the work in \cite{kim2015velocity} that extended velocity-based collision avoidance models by incorporating contact, pushing, and resistive forces to represent balance recovery strategies. 
Despite their differences, microscopic pedestrian models for dense crowds remain grounded in simplified two-dimensional, disk-like, representations of individual agents.
Beyond this simple isotropic approach, submicroscopic body representation have also been proposed. In \cite{Thompson_1995}, a representation featuring three circles for each agents is used in order to account for pedestrians shoulders. This approach have recently been augmented to proposed multi-circles agent body shapes fitted on an anthropometric database \cite{Dufour_2025}. Humanoid limbed representation have also recently been proposed to simulate pedestrians. However, this approach uses limited physical interaction modelling without consideration for balance recovery mechanism, hence similar to former approaches \cite{Shang_2025}.
The study \cite{wang2019modeling} addresses cascading loss of balance following external perturbations—known as the \emph{domino effect}. However, this approach remains restricted to perturbation propagation along a one-dimensional line of individuals and cannot, in its current form, be applied for dense crowd modelling.

None of the existing refined modelling paradigms for dense crowds explicitly incorporate balance-control mechanisms arising from the coupled motion of the upper body and the legs during physical interactions. 
In fact, current dense-crowd models largely neglect the biomechanical coupling that governs how pedestrians maintain balance under contact. 
However, this coupling is essential for reproducing large-scale instabilities and emergent collective patterns driven by balance-control dynamics. 
Rather than further increasing the complexity of existing microscopic models through ad hoc contact forces, we introduce in this study a original dense-crowd model grounded in minimalist biomechanical principles, which explicitly accounts for the coupling between the legs and the upper body during physical interactions. 
Numerical simulations demonstrate that the model naturally reproduces several empirically observed collective phenomena within an interpretable submicroscopic framework.
The model formulation and parameter interpretation are presented in the next section. Section~\ref{Simulation} reports the simulation results, including different collective dynamical regimes and the corresponding phase diagram. Conclusions and possible extensions of the model are discussed in Section~\ref{Conclusion}.

\section{Two-level pedestrian model\label{ModelDefinition}}

\begin{figure}[!ht]
    \bigskip
    \input{Figs/2LevelScheme}
    \centering
    \caption{Scheme for the two-level pedestrian model. The force $\mathbf u$ acting on the upper body is an unbalancing mechanism whereby vertical displacements between the body and the legs induce instability due to the gravity. Controversially, the force $\mathbf b$ displays a balancing mechanism whereby the legs relax to the upper body to maintain the body’s upright posture.}
    \label{fig:2LPM}
\end{figure}
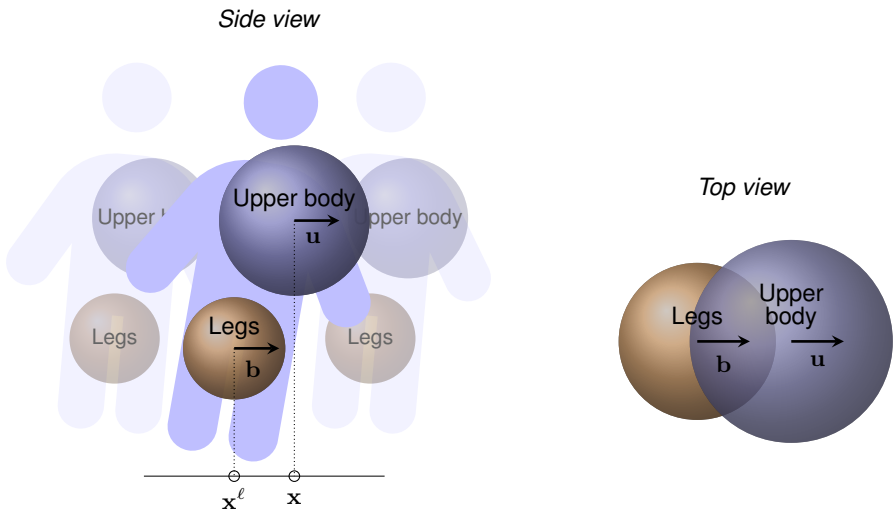

In this work, our main hypothesis is that the loss and the recovery of standing balance are dominant factors explaining the mechanisms of individual absorption and transmission of perturbation forces, and therefore explaining the collective motion of ultra-dense crowds. To test this hypothesis, we introduce a minimal two-level pedestrian model in which each agent consists of a upper body and a leg subsystem. 
We consider the relative positions, in the horizontal plane, of the upper body on the one hand and the legs on the other (see Fig.~\ref{fig:2LPM}). These positions, denoted in the following $\mathbf x_n(t)$ for the upper body and $\mathbf x_n^\ell(t)$ for the legs of the $n$-th pedestrian at time $t$, can be interpreted respectively as the ground-projected position of the body’s center of mass and the center position of the support area containing the feet. The interaction between the upper body and the legs relies on two antagonistic mechanisms:
\begin{itemize}
    \item First, when the upper body position moves away from the legs, it become affected by a destabilizing force. This increases the relative velocity of the upper body with respect to the legs and further amplifies this displacement. This dynamic reflects the fact that a upper body displaced from its support base naturally tends to ``fall''. In other words, this force could be seen as representing the linearized effect of gravity on pedestrians upper bodies. This \emph{unbalancing mechanism}, unintentionally endured by the pedestrian, is modelled by the term $\mathbf u$ in Fig.~\ref{fig:2LPM}.
    \item Second, in response, the legs move to ensure a stabilization role: a rebalancing force increases their velocity to bring them back under the upper body, compensating for the destabilizing dynamics. This \emph{balancing mechanism}, intentionally produced by the pedestrian, corresponds to the term $\mathbf b$ in Fig.~\ref{fig:2LPM}.
\end{itemize}
These mechanisms dynamically couple the upper body and leg layers, forming a quasi-three-dimensional paradigm that integrates vertical and horizontal interactions without explicitly simulating full 3D body dynamics.

Beside the kinematic relationships between projected positions and velocities of the pedestrian bodies and legs, namely $\dot {\mathbf x}_n(t)={\mathbf v}_n(t)$ and $\dot {\mathbf x}^\ell_n(t)={\mathbf v}^\ell_n(t)$, the dynamics of the two-level pedestrian model are given by 
\begin{equation}
        \dot {\mathbf v}_n(t)=\underbrace{\lambda_u\big({\mathbf u}_n(t) - {\mathbf v}_n(t)\big)}_{\text{Unbalancing}}-\underbrace{\lambda {\mathbf v}_n(t)}_{\text{Damping}}-\underbrace{\sum_{m\ne n}\nabla V\big({\mathbf x}_n(t)-{\mathbf x}_m(t)\big)}_{\text{Upper body interaction}}
    \label{eq:2LPMa}
\end{equation}
for the upper body and
\begin{equation}
\dot {\mathbf v}^\ell_n(t)=\underbrace{\lambda_b\big({\mathbf b}_n(t) - {\mathbf v}^\ell_n(t)\big)}_{\text{Balance recovery}}-\underbrace{\sum_{m\ne n}\nabla V_\ell\big({\mathbf x}^\ell_n(t)-{\mathbf x}^\ell_m(t)\big)}
        _{\text{Legs interaction}}
    \label{eq:2LPMb}
\end{equation}
for the legs, where 
\begin{equation}
    {\mathbf b}_n(t)=v_b(t)\,{\mathbf e}_n(t)
\end{equation} is the intentional balance recovery force acting on the legs with $v_b\ge0$ the balance recovery speed, 
while 
\begin{equation}
    {\mathbf u}_n(t)=v_u(t)\,{\mathbf e}_n(t)
\end{equation} is the unintentionally endured unbalancing force acting on the upper body with $v_u\ge0$ the unbalancing speed.
Here $\mathbf e_n(t)=\big({\mathbf x}_n(t)-{\mathbf x}^\ell_n(t)\big)/\big|{\mathbf x}_n(t)-{\mathbf x}^\ell_n(t)\big|$ is the unit vector that points from the legs to the upper body.
In this two–sublevel model, three types of behaviour are distinguished:
\begin{enumerate}
    \item[(a)] \textbf{Balance/unbalance behaviour: Upper body and leg coupling}~ For simplicity, both the balancing and unbalancing processes are modelled using linear relaxation, which remains dominated in magnitude by the repulsive interaction forces at short distances. 
    These mechanisms are controlled by the balancing and unbalancing rates $\lambda_b$ and $\lambda_u$. In addition, the balance and unbalance speeds, $v_b$ and $v_u$, are assumed to be constant and equal. In principle, they could differ or dynamically depend on the distance between legs and upper body or their respective velocities.

    \item[(b)] \textbf{Interaction behaviour: Two-layer upper body and leg repulsions}~ We assume that upper bodies and legs interact independently on the two layers through purely repulsive, radially symmetric potentials $V$ and $V_\ell$, which monotonically vanish with increasing distance. These pairwise interaction forces describe repulsion and exclusion between the bodies and the legs respectively. They are nonlinear and dominate the upper body/leg coupling mechanisms at short distance.
    In the following, the interaction potentials are the exponential (short-range) functions \cite{helbing1995social} 
    \begin{equation}
    V(x)=AB\exp\left(\frac{-|x|}{B}\right)\qquad\text{and}\qquad V_\ell(x)=AB_\ell\exp\left(\frac{-|x|}{B_\ell}\right),\qquad A,B,B_\ell>0.
    \end{equation} 
    Other repulsive interaction potentials, such as algebraic and power-law repulsion potentials as in \cite{silverberg2013collective,bottinelli2016emergent}, can be used instead. In fact, the collective dynamics of the model are mainly governed by the balance/unbalance mechanisms and related parameters, provided that the interaction potentials are nonlinear and repulsive.
    Because the pedestrian upper body is wider than the ground support area, we assume a larger repulsive distance at upper body level than at leg level, i.e. $B_\ell< B$, which naturally makes the legs more mobile and able to adjust more freely under dense-contact conditions.

    \item[(c)] \textbf{Damping: Body resistance to external perturbation}~ The damping term of the upper body $-\lambda v_n$, where $\lambda>0$ is the damping rate, represent pedestrian's active resistance to external perturbations. 
    This term may include an external desired velocity $v^*$ via the relaxation term $-\lambda (v_n-v^*)$, to provide a direction of motion to the pedestrians .
\end{enumerate}

Having defined the interactions, it is instructive to clarify the qualitative behaviour expected from such a system. Each pedestrian can be viewed as a mechanically coupled body–leg unit, analogous to an inverted pendulum: the upper body provides inertia and is inherently destabilized by gravity when displaced from its support base, while the legs act as an active control element that attempts to restore equilibrium. 
The inverted pendulum is widely used as a conceptual model of individual pedestrian balance control \cite{winter1998stiffness,peterka2002sensorimotor}.
When many such units interact through short-range exclusion forces, perturbations experienced by one pedestrian are transmitted mechanically to others thanks to the interaction layers, allowing local balance corrections or instabilities to propagate through the crowd. From this perspective, the model constitutes a many-body system of interacting inverted pendula, where energy injected by imbalance and dissipated through balance recovery is continuously exchanged between neighbouring pedestrians. The explicit two-layer structure further introduces internal degrees of freedom that naturally favour oscillatory responses and feedback loops, creating the conditions for collective modes to emerge at the system scale. These considerations motivate a systematic exploration of the model’s dynamical regimes in simulation, where macroscopic organization can arise from the interplay between individual balance control and mechanical coupling.

\section{Simulation results \label{Simulation}}

The duality between stabilizing and destabilizing forces generates a rich repertoire of individual behaviours, giving rise not only to crystallization but also to density waves and collective chiral motion. 
Beyond visualising the evolution of pedestrian upper body and leg positions over time, two main order parameters are used to quantify the collective behaviour:
\begin{itemize}
\item \emph{System kinetic energy}
\begin{equation}
\mathcal E = \sum_{n=1}^N |\mathbf v_n|^2 \ge 0.
\end{equation}
The kinetic energy $\mathcal E$ characterizes the \emph{magnitude of fluctuations} in pedestrian velocities.
\item \emph{Local velocity correlation} 
\begin{equation}
    \mathcal C = \frac{1}{N} \sum_{n=1}^N \frac{1}{N_n} \sum_{m \in D_n} \frac{\mathbf v_n \cdot \mathbf v_m}{|\mathbf v_n||\mathbf v_m|} \in [-1,1],
\end{equation}  
where $D_n = \{ m,\ |\mathbf x_n - \mathbf x_m| < r \}$ is the set of neighbouring pedestrians within a radius $r = 2$~m, and $N_n = \text{card}(D_n)$ is the number of neighbours.  
The velocity correlation $\mathcal C$ quantifies the \emph{local alignment of motion}, indicating how coherently pedestrians move together.
\end{itemize}
In the following, we consider a square periodic system and two sets of parameters. 
The first set, presented in Section~\ref{Wave}, promotes the formation of density waves, while the second set, presented in Section~\ref{Chiral}, gives rise to collective chiral oscillations and coherent rotational motion.  
Finally, the phase diagram of the model as a function of the balance and unbalance parameters is presented in Section~\ref{PhaseDiagram}.
Details on the simulation setup and numerical scheme as given in the Supplementary  material.

\subsection{Density wave \label{Wave}}

\begin{figure}[!ht]
\def\lf{.30\textwidth}
    \centering
   \includegraphics[width=\lf]{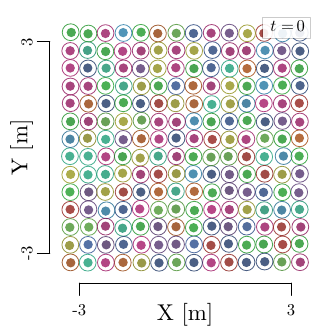}\hspace{10mm}\includegraphics[width=\lf]{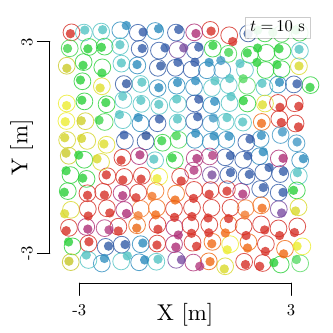}\\[3mm]
   \includegraphics[width=\lf]{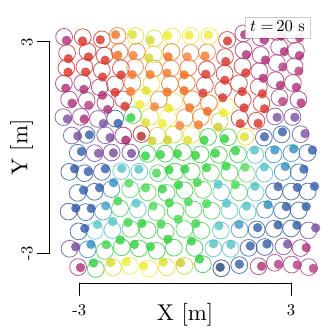}\hfill
   \includegraphics[width=\lf]{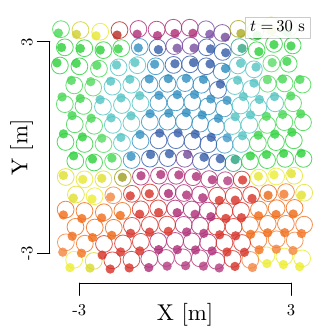}\hfill\includegraphics[width=\lf]{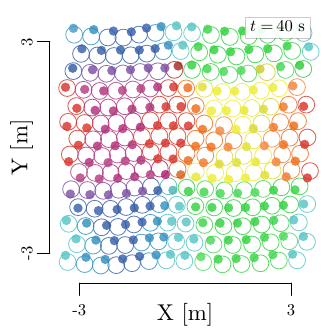}
    \caption{Time evolution of pedestrian body and leg positions under the density-wave parameter setting. A diagonal density wave rapidly emerges and propagates through the system. Colour encodes pedestrian direction, and brightness indicates velocity magnitude.}
    \label{fig:Positions1}
\end{figure}

When the unbalancing rate and the balance/unbalance speed are set to intermediate values, the system exhibits a density-wave pattern. 
Pedestrians move collectively, forming coherent diagonal waves across the domain, while maintaining local alignment. 
These waves seem qualitatively closed to the self-sustained waves observed during Oasis concert \cite{bottinelli2018can}.
The time evolution of pedestrian upper body and leg positions at discrete times $t=0,10,20,30$ and $40$~s is shown in Figure~\ref{fig:Positions1}, illustrating the emergence and propagation of stationary density wave propagating in diagonal. 
The parameter values are the following: $\lambda_u=0.5$~s$^{-1}$, $\lambda_b=1$~s$^{-1}$, and $v=1$~m/s. 
The values of the remaining parameters are given in the Supplementary material.

\begin{figure}[!ht]
    \centering\bigskip
    \includegraphics[height=5cm]{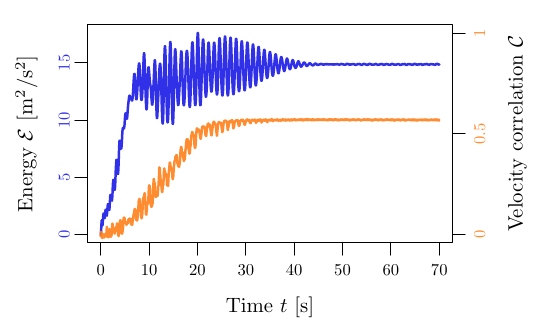}\hfill
    \includegraphics[height=5cm]{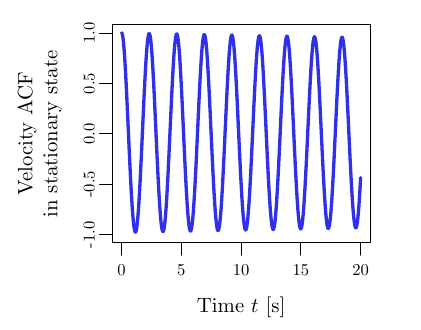}\vspace{-2.5mm}
    \caption{System kinetic energy and velocity correlation time-series (left panel) corresponding to the sequence shown in Figure.~\protect\ref{fig:Positions1}, and velocity time autocorrelation function (ACF) in stationary state (left panel). The system converges to a stationary state featuring a density wave and pedestrians exhibiting periodic velocity fluctuations at high frequency.}
    \label{fig:EnergyACF1}
\end{figure}

The corresponding time series of system kinetic energy and velocity correlation (left panel) and velocity autocorrelation in the stationary state (right panel) are shown in Figure~\ref{fig:EnergyACF1}. 
Across the density-wave domain, the velocity correlation remains approximately constant, indicating coherent motion, while the kinetic energy increases with the unbalancing rate and speed, reflecting waves of varying amplitude. 
The full video of the density-wave experiment is available in the Supplementary Material.

\newpage
\subsection{Chiral oscillation \label{Chiral}}

\begin{figure}[!ht]
\def\lf{.30\textwidth}
\bigskip
    \centering
   \includegraphics[width=\lf]{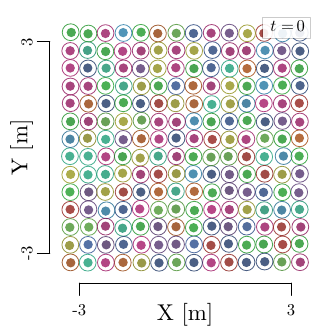}\hspace{10mm}\includegraphics[width=\lf]{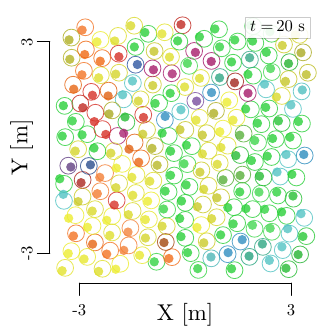}\\[3mm]
   \includegraphics[width=\lf]{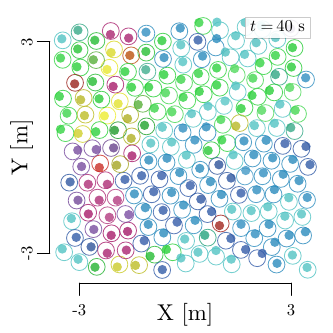}\hfill
   \includegraphics[width=\lf]{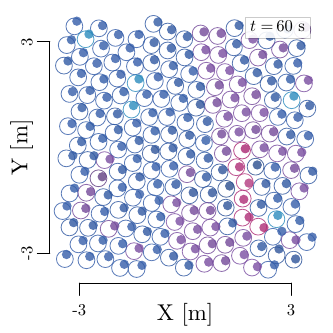}\hfill
   \includegraphics[width=\lf]{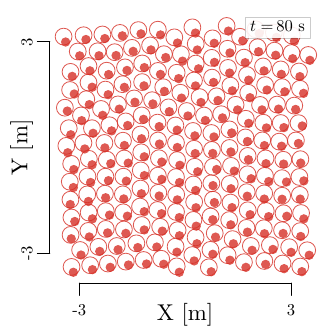}
    \caption{Time evolution of pedestrian body and leg positions under the chiral-oscillation parameter setting. Pedestrians self-organize into a collective rotational motion. Color encodes pedestrian direction, and brightness represents velocity magnitude. 
    }
    \label{fig:Positions2}
\end{figure}

\begin{figure}[!ht]
    \centering\bigskip
    \includegraphics[height=5cm]{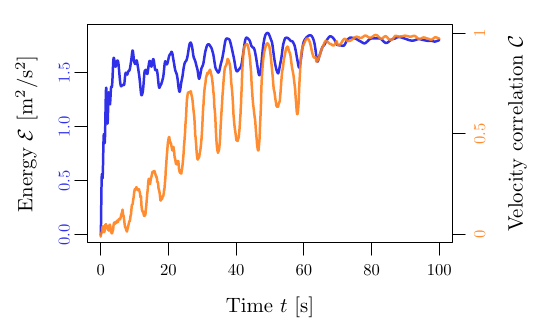}\hfill
    \includegraphics[height=5cm]{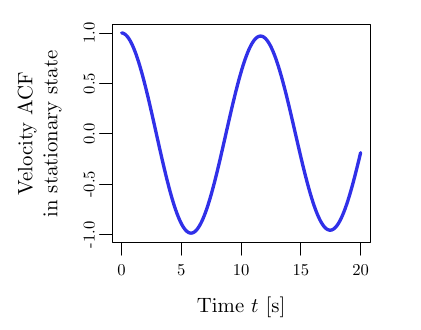}\vspace{-2.5mm}
    \caption{System kinetic energy and velocity correlation time-series (left panel) corresponding to the sequence shown in Figure.~\protect\ref{fig:Positions2}, and velocity time autocorrelation function (ACF) in stationary state (left panel). The system converges to a stationary state exhibiting a global collective behavior with chiral oscillations, where pedestrians display periodic velocity fluctuations at low frequency.}
    \label{fig:EnergyACF2}
\end{figure}

Under conditions where the unbalancing rate exceeds the balancing rate and the balance/unbalance speed is low, the system exhibits a global collective behaviour characterized by chiral oscillations. 
In this regime, pedestrians move coherently, describing individual circular trajectories while maintaining synchronized motion -- a behaviour that has been documented in real crowds, notably at the St. Fermín festival \cite{gu2025emergence}. 
The time evolution of pedestrian positions at discrete times $t=0,20,40,60$ and 
$80$~s is illustrated in Figure~\ref{fig:Positions2}, while the corresponding dynamics of system kinetic energy, velocity correlation, and velocity autocorrelation are shown in Figure~\ref{fig:EnergyACF2}. 
The parameter values are: $\lambda_u=1$~s$^{-1}$, $\lambda_b=0.5$~s$^{-1}$, and $v=0.2$~m/s. 

After a short simulation time, pedestrians walk in circles at a frequency of around 12 seconds per cycle (see Fig.~\ref{fig:EnergyACF2}, right panel). 
Their velocity varies synchronously and their spatial correlation is close to one (see Fig.~\ref{fig:EnergyACF2}, left panel).  
Surprisingly, even minimal local coupling between legs and body can generate large-scale, coherent rotational dynamics observed empirically in dense pedestrian systems. 
In addition, the amplitude of these rotations can be partly tuned by adjusting the balance/unbalance speed $v$.
The full video of the chiral oscillation experiment is available in the Supplementary Material.

\newpage
\subsection{Phase diagram \label{PhaseDiagram}}

\begin{figure}[!ht]
    \centering\bigskip
    \includegraphics[width=0.27\textwidth]{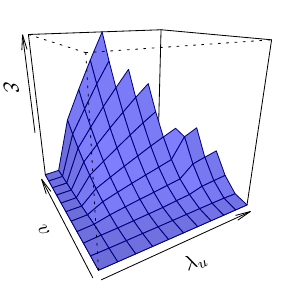}\hspace{37mm}
    \includegraphics[width=0.27\textwidth]{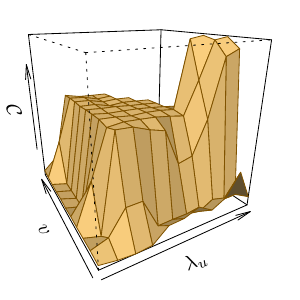}\\[2mm]
    \includegraphics[width=0.48\textwidth]{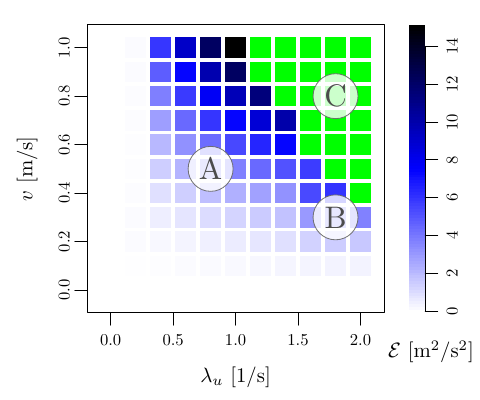}\hfill
    \includegraphics[width=0.48\textwidth]{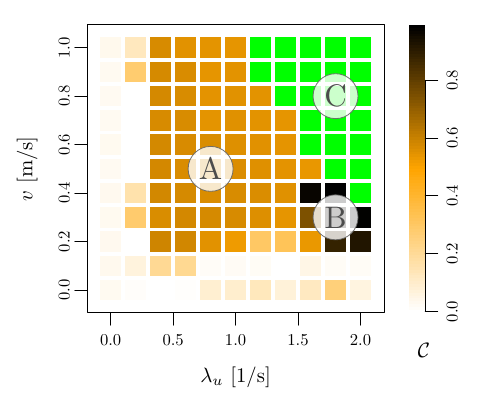}\vspace{0mm}
    \caption{Phase diagram of the two-level pedestrian model as a function of unbalancing rate $\lambda_u$ and balance/unbalance speed $v = v_u = v_b$ ($\lambda_b = 1$~s$^{-1}$). Four main states emerge with increasing $\lambda_u$ and $v$: a crystallization state with low energy and correlation; (A) a density wave state at relatively low unbalancing rate with intermediate correlation; (B) a chiral oscillating state at high unbalancing rate with high correlation; and (C) a disordered state where leg and body behaviours are incoherent (shown in green).}
    \label{fig:PhaseDiagram}
\end{figure}

We systematically explore the dynamics of the two-level pedestrian model by varying the unbalancing rate $\lambda_u$ and the balance/unbalance speed $v = v_u = v_b$, while keeping the balancing rate fixed at $\lambda_b = 1$~s$^{-1}$. 
Figure~\ref{fig:PhaseDiagram} shows the system kinetic energy (left panels) and local velocity correlation (right panels) in the stationary state, used to characterize the different system phases.
The kinetic energy quantifies the amplitude of pedestrian motion, while the velocity correlation measures the degree of coherence among pedestrians. In the crystallization state, energy and correlation are both low. In the density-wave state, energy is intermediate and correlation is high, reflecting coherent wave propagation. In the chiral-oscillation state, energy remains low while correlation is high, corresponding to slow, coordinated rotational motion.

As either $\lambda_u$ or $v$ cross specific thresholds, the system transitions from a crystallized state to a density-wave pattern, forming a diagonal boundary beyond which the dynamics become disordered. 
In addition, in the bottom-right region of the parameter space, where $\lambda_u$ is high and $v$ is low, the system exhibits collective chiral oscillations. 
Notably, the velocity correlation remains approximately constant across the entire density-wave domain, whereas the system kinetic energy increases with both $\lambda_u$ and $v$. 
This observation indicates that the density waves share similar coherence and characteristic propagation speed and frequency. 
However, their amplitudes vary depending on the specific parameter values. 
This highlights the tunability of collective wave behaviour through the two-level pedestrian interactions.
In addition, further simulation results show that similar collective dynamics of density waves and chiral oscillations arise when using the algebraic repulsive interaction potentials
\begin{equation}
    V(x)=\frac{AB}{|x|}\qquad\text{or}\qquad V(x)=AB\bigg(1-\frac{|x|}B\bigg)_+^{5/2}
\end{equation}
where $A,B>0$. In fact, the main mechanisms governing the emergence of collective motions are the balance and unbalance dynamics, and related parameters $\lambda_b$ and $\lambda_u$. %, $v_b$ and $v_u$. 
The interaction potentials appear to be secondary, as long as they are repulsive. 
Collective dynamics also emerge in the absence of potential leg interactions $V_\ell$, although synchronisation is more difficult to achieve in this case.
Furthermore, emerging dynamics appear to be relatively insensitive to the damping rate $\lambda$, provided that it is strictly positive.

\section{Discussion \label{Conclusion}}

Although deliberately minimalist, the present two-level model reveals a qualitative advance with respect to existing approaches to dense crowd dynamics. 
Classical two-dimensional models based solely on contact forces or velocity alignment successfully reproduce congestion, jamming, or stop-and-go motion, but they fail so far to account, within a single and interpretable framework, for the coexistence of distinct instability-driven collective modes that are directly relevant for safety of dense crowds. 
In contrast, by explicitly introducing an internal body–leg degree of freedom and balance-related feedback at the individual scale, the present model is able to reproduce propagating density waves and coherent chiral oscillations, emerging from purely local interactions.
Importantly, these behaviours arise without fine-tuning or ad hoc global rules.  
Systematic exploration of the parameter space reveals well-structured dynamical regimes and robust phase boundaries, indicating that the observed collective states are intrinsic characteristics of the model. 
This robustness, together with the minimal and physically interpretable nature of the balance and unbalance mechanisms, strongly supports the relevance of biomechanical coupling as a missing ingredient in previous dense crowd models. In addition, the precise functional form of the repulsive interaction potentials appears to be secondary for the emergence of collective dynamics, provided that interactions remain short-ranged, repulsive and offer more mobility to the legs. 
This further emphasizes that the dominant organizing principle lies in the internal balance dynamics rather than in the details of contact modelling.

At a mechanical level, the two-level pedestrian representation can be interpreted as a system of interacting inverted pendulums. 
Each pedestrian consists of an upper body acting as a mass subject to gravity-driven destabilization, supported by a leg subsystem that plays the role of an actively controlled base providing elastic and dissipative stabilization. The relative displacement between legs and upper body defines a local tilt, so that the unbalancing term mimics the linearized gravitational torque of an inverted pendulum, while the balancing term represents active postural control that repositions the support base to maintain equilibrium. 
This representation is classical to model individual pedestrian posture control \cite{winter1998stiffness,peterka2002sensorimotor}. 
When embedded in a dense crowd, the individual inverted pendulums are mechanically coupled through short-range repulsive interactions, allowing energy, momentum, and perturbations to be transmitted across the system. The crowd can thus be viewed as a spatially extended network of coupled inverted pendulums or mass–spring–damper units, whose collective dynamics are governed by both internal balance control and inter-agent mechanical (exclusion) constraints.
This mechanical interpretation naturally connects the balance–unbalance feedback to the theory of coupled oscillatory systems, multi-layer systems, and Kuramoto-type systems 
Within this framework, stability and synchronization theory \cite{Kuramoto2003,acebron2005kuramoto,redig2023ergodic,wang2025phase} provides a powerful conceptual lens to understand how local balance controls can self-organize into macroscopic patterns and system-wide coordinated motion.

Nonetheless, important steps remain before the model can serve as a predictive tool: it must be extended with non-linear coupling terms and state-dependent parameters (e.g., making relaxation rates depend on body–leg separation or relative velocities), generalized to heterogeneous anisotropic populations with individual-specific biomechanical parameters, and augmented with frictional forces and rupture/falling mechanisms that capture discontinuous loss-of-equilibrium events. 
Crucially, beyond model extensions, systematic calibration and validation against experimental and naturalistic datasets (video tracking from concerts, festivals, and pilgrimage flows, and controlled lab experiments) are required to estimate parameters, test predictive skill, and evaluate safety-relevant diagnostics. 
These developments will determine whether the two-level paradigm can be elevated from a qualitative mechanistic framework to a quantitatively reliable model for forecasting and mitigating hazards in ultra-dense crowds.

\paragraph{Acknowledgments}
The authors acknowledge the use of ChatGPT (OpenAI, GPT-5 Mini) and DeepL Translator (DeepL GmbH) for assistance in refining the English language of the manuscript. 
All scientific ideas, analyses, and interpretations reported in this work are solely those of the authors.

\paragraph{Author contributions statement}
TC: conceptualization, formal analysis, investigation, writing (review \& editing).
MC: conceptualization, writing (review \& editing), supervision. 
JP: conceptualization, writing (review \& editing), supervision. 
AS: conceptualization, writing (review \& editing), supervision. 
AT: conceptualization, methodology, formal analysis, investigation, writing (original draft).

\bibliographystyle{abbrv}
\bibliography{refs}

\newpage

\section*{Supplementary material \label{SupMat}}

\subsection*{Simulation setup}

We consider in the simulations a square system of 7$\times$7 meters with periodic boundaries (i.e., a torus) and 196 pedestrians. 
The model equation \eqref{eq:2LPD} is simulated using the numerical scheme
\def\dt{\delta t}
\begin{equation}
        \left\{~~\begin{aligned}
        &{\mathbf x}_n(t+\dt)={\mathbf x}_n(t)+\dt{\mathbf v}_n(t+\dt)\\[1mm]
        &{\mathbf x}^\ell_n(t+\dt)={\mathbf x}^\ell_n(t)+\dt{\mathbf v}^\ell_n(t+\dt)\\[1mm]
        &{\mathbf v}_n(t+\dt)={\mathbf v}_n(t)+\dt\bigg[\lambda_u\big(v{\mathbf e}_n(t) - {\mathbf v}_n(t)\big)-\lambda {\mathbf v}_n(t)-\sum_{m\ne n}\nabla V\big({\mathbf x}_n(t)-{\mathbf x}_m(t)\big)\bigg]\\[1mm]
        &{\mathbf v}^\ell_n(t+\dt)={\mathbf v}^\ell_n(t)+\dt\bigg[\lambda_b\big(v{\mathbf e}_n(t) - {\mathbf v}^\ell_n(t)\big)-\sum_{m\ne n}\nabla V_\ell\big({\mathbf x}^\ell_n(t)-{\mathbf x}^\ell_m(t)\big)\bigg]
    \end{aligned}\right.
    \label{eq:2LPD}
\end{equation}
where the position of the legs and upper body are computed using an implicit Euler scheme while the velocity are computed using an explicit scheme.
The repulsive potentials for upper bodies and legs, $V$ and $V_\ell$, are the exponential functions \cite{helbing1995social}
\begin{equation}
    V(x)=AB\exp\left(\frac{-|x|}{B}\right)\qquad\text{and}\qquad V_\ell(x)=AB_\ell\exp\left(\frac{-|x|}{B_\ell}\right).
\end{equation}
The parameter values are the following. 
The damping rate $\lambda$ is set to 1~s$^{-1}$. 
The repulsion rate $A$ is set to 5~m/s$^2$ for both leg and body repulsive potential while the characteristic repulsion distance is higher for the body, $B=0.5$~m, than for the legs, $B_\ell=0.3$~m. 
For the sake of simplicity, the balance and imbalance speed are assumed to be constant and equal to $v$. 
The values of the parameters belonging to the two-level pedestrian model are given in Table~\ref{tab:ParaValue}.

\begin{table}[!ht]
    \centering
    \renewcommand{\arraystretch}{1.7} 
    \begin{tabular}{l|c|c|c}
        &$\lambda_b$ [s$^{-1}$]&$\lambda_u$ [s$^{-1}$]&$v$ [m/s]\\
        \hline
        Density wave -- Sec.~\ref{Wave}, Figs~\ref{fig:Positions1} and \ref{fig:EnergyACF1}&1&0.5&1\\[-2mm]
        Chiral oscillation -- Sec.~\ref{Chiral}, Figs~\ref{fig:Positions2} and \ref{fig:EnergyACF2}$\quad$&0.5&1&0.2\\[-2mm]
        Phase diagram -- ~Sec.~\ref{PhaseDiagram}, Figs~\ref{fig:PhaseDiagram}&1&0 $\to$ 2&0 $\to$ 1
    \end{tabular}
    \caption{Setting of the parameters own to the two-level pedestrian in Section~\ref{Wave}, see Figures~\ref{fig:Positions1} and \ref{fig:EnergyACF1}, Section~\ref{Chiral}, see Figures~\ref{fig:Positions2} and \ref{fig:EnergyACF2}, and Section~\ref{PhaseDiagram}, see Figure~\ref{fig:PhaseDiagram}.}
    \label{tab:ParaValue}
\end{table}

The time step is set to 0.01~s in Figures ~\ref{fig:Positions1} and \ref{fig:EnergyACF1}, \ref{fig:Positions2}, and \ref{fig:EnergyACF2} while $\delta t$ is set to 0.1~s in the simulations of the phase diagram in Figure~\ref{fig:PhaseDiagram}. 
In addition, the system performance in stationary states shown in  Figures~\ref{fig:EnergyACF1} and \ref{fig:EnergyACF2}, right panels, and in  Figure~\ref{fig:PhaseDiagram} is obtained after a simulation time of 500~s and 1000~s, respectively. 
In addition, 1000 attempts are carried out before to state that the system is disordered with incoherent leg and body behaviours.
The initial conditions are identical across simulations, consisting of an upright square lattice perturbed by independent, normally distributed noise with an amplitude of 1 cm.

\newpage

\subsection*{Videos}
The videos of the simulation for the wave formation shown in Section~\ref{Wave}, see Figures~\ref{fig:Positions1} and \ref{fig:EnergyACF1}, and for the formation of chiral oscillations in Section~\ref{Chiral}, see Figures~\ref{fig:Positions2} and \ref{fig:EnergyACF2}, are available at
\begin{center}
    \href{https://uni-wuppertal.sciebo.de/s/3e2spF9ZLJL2JEJ}{\texttt{https://uni-wuppertal.sciebo.de/s/3e2spF9ZLJL2JEJ}}
\end{center}

\subsection*{Online simulation module}
An online simulation module of a reduced square system of 5$\times$5 meters with periodic boundaries and 100 pedestrians is available at
\begin{center}
    \href{https://antoinetordeux.github.io/Two-Level-Pedestrian-Model/}{\texttt{https://antoinetordeux.github.io/Two-Level-Pedestrian-Model/}}
\end{center}

\end{document}

%% file: Figs/2LevelScheme.tex
\centering
\vspace{5mm}
\begin{minipage}[b]{\textwidth}
    
\hspace{10mm}
\textcolor{white!95!blue}{\hspace{-5mm}
\scalebox{14}{\rotatebox{-5}{\faPerson}}\hspace{-2mm}
\scalebox{14}{\rotatebox{-5}{\faPerson}}}\\[-38mm]
\begin{tikzpicture}
\hspace{17mm}
\fill[ball color=orange, opacity=0.15] (.8,0) circle (6mm);
\node at (.8,0) {\textcolor{white!40!black}{\sffamily\footnotesize Legs}};
\fill[ball color=orange, opacity=0.15] (4.2,0) circle (6mm);
\node at (4.2,0) {\textcolor{white!40!black}{\sffamily\footnotesize Legs}};
\fill[ball color=blue, opacity=0.1] (1.3,1.6) circle (8mm);
\node at (1.3,1.6) {\textcolor{white!40!black}{\sffamily\footnotesize Upper body}};
\fill[ball color=blue, opacity=0.1] (4.7,1.6) circle (8mm);
\node at (4.7,1.6) {\textcolor{white!40!black}{\sffamily\footnotesize Upper body}};
\end{tikzpicture}\vspace{-53mm}
\begin{center}
\begin{minipage}[b]{.3\textwidth}
\centering {\sffamily\small \textit{~~~~~Side view}$\quad$}\\[2mm]
\textcolor{white!75!blue}{\hspace{-3mm}\scalebox{15}{\rotatebox{-10}{\faPerson}}}\\[-44.5mm]
\begin{tikzpicture}[>=stealth, font=\sffamily]
  \draw[-, line width=.25pt] (0,0) -- (3.2,0);
  \coordinate (P1) at (1.2,1.7);
  \fill[ball color=orange!50!white, opacity=0.75] (P1) circle (6.8mm);
  \node[above=0pt] at (P1) {{\small Legs}};
  \coordinate (P2) at (2,3.4);
  \fill[ball color=blue!30!white, opacity=0.75] (P2) circle (10mm);
  \node[above=0pt] at (P2) {{\small Upper body}};

  \draw[densely dotted] (P1) -- (1.2,0);
  \draw[densely dotted] (P2) -- (2,0);
  
  %\draw[->, line width=1pt] (1.2,0) -- (1.8,0) node[anchor=west] {};
  \draw[->, line width=1pt] (P1) -- (1.8,1.7) node[anchor=west] {};
  \node[below=1pt] at (1.45,1.7) {\small$\mathbf b$}; 
  \draw[->, line width=1pt] (P2) -- (2.6,3.4) node[anchor=west] {};
  \node[below=1pt] at (2.25,3.4) {\small$\mathbf u$};

  \draw (1.2,0) circle (2pt) node[below=1pt] {\small$\mathbf x^\ell$};
  \draw (2,0) circle (2pt) node[below=3pt] {\small$\mathbf x$};
\end{tikzpicture}
\end{minipage}\hspace{20mm}\begin{minipage}[b]{.3\textwidth}
\centering {\sffamily\small \textit{Top view}$\quad$}\\[5mm]
\begin{tikzpicture}[>=stealth, font=\sffamily]
  % \coordinate (P1) at (1,2.5);
  % \fill[ball color=orange!50!white, opacity=0.75] (P1) circle (10.45mm);
  % \node[above=1pt] at (P1) {{\small Legs}};
  % \coordinate (P2) at (2,1.5);
  % \fill[ball color=blue!30!white, opacity=0.75] (P2) circle (13.5mm);
  % \node[above=-3pt] at (P2) {{\small Upper body}};

  % \draw[->, line width=1pt] (P1) -- (1.5,2) node[anchor=west] {};
  % \node[below=1.8pt] at (1.05,2.4) {\small$\mathbf b$}; 
  % \draw[->, line width=1pt] (P2) -- (2.5,1) node[anchor=west] {};
  % \node[below=1.8pt] at (2.05,1.35) {\small$\mathbf u$};
  \coordinate (P1) at (1,2.5);
  \fill[ball color=orange!50!white, opacity=0.75] (P1) circle (10.45mm);
  \coordinate (P2) at (2.25,2.5);
  \fill[ball color=blue!30!white, opacity=0.6] (P2) circle (13.5mm);
  \node[above=1pt] at (P1) {{\small Legs}};
  \node[above=10pt] at (P2) {{\small Upper}};
  \node[above=1pt] at (P2) {{\small body}};

  \draw[->, line width=1pt] (P1) -- (1.7,2.5) node[anchor=west] {};
  \node[below=1.8pt] at (1.35,2.5) {\small$\mathbf b$}; 
  \draw[->, line width=1pt] (P2) -- (2.95,2.5) node[anchor=west] {};
  \node[below=1.8pt] at (2.6,2.5) {\small$\mathbf u$};
\end{tikzpicture}\vspace{10mm}
\end{minipage}\vspace{-2mm}
\end{center}
\end{minipage}